\documentclass[10pt,3p]{elsarticle}
\usepackage{graphicx}
\usepackage{float}
\usepackage{subfig}
\usepackage{setspace}
\usepackage{color}
\usepackage{multirow}
\usepackage{amsmath}

\graphicspath{./Figures}
\DeclareMathOperator{\Tr}{Tr}

\begin{document}
\title{Upstream open loop control of the recirculation area downstream of a backward-facing step}
\author{N. Gautier and J.-L. Aider}
\address{Laboratoire de Physique et M\'ecanique des Milieux H\'et\'erog\`enes (PMMH), UMR7636 CNRS, \\ \'Ecole Sup\'erieure de Physique et Chimie Industrielles de la ville de Paris\\ 10 rue Vauquelin,  75005 Paris, France}

\begin{abstract}
The flow downstream a backward-facing step is controlled using a pulsed jet placed upstream of the step edge. Experimental velocity fields are computed and used to the recirculation area quantify. The effects of jet amplitude, frequency and duty cycle on this recirculation area are investigated for two Reynolds numbers ($Re_h = 2070$ and $Re_h=2900$). The results of this experimental study demonstrate that upstream actuation can be as efficient as actuation at the step edge when exciting the shear layer at its natural frequency. Moreover it is shown that it is possible to minimize both jet amplitude and duty cycle and still achieve optimal efficiency. With minimal amplitude and a duty-cycle as low as 10\% the recirculation area is nearly canceled.
\end{abstract}


\maketitle

\section{Introduction}
Separated flows are ubiquitous in nature and industrial processes. They occur in many devices such as combustion chambers, air conditioning plants, moving ground and air vehicles (see \cite{CHC,Hucho2005,Weisenstein2000}). The main feature of separated flows is the recirculation bubble i.e. the region where the direction of the flow is reversed \cite{Armaly1983}. In most industrial applications it is important to reduce recirculation in order to improve drag, increase lift, suppress vibrations or lower aeroacoustic noise. Sometimes an increase in recirculation is welcome, for instance to increase mixing in a combustion chamber.
\\\\

\begin{figure}[H]
\centering
\subfloat[Sketch of the BFS geometry and definition of the main parameters]{\includegraphics[width=0.45\textwidth]{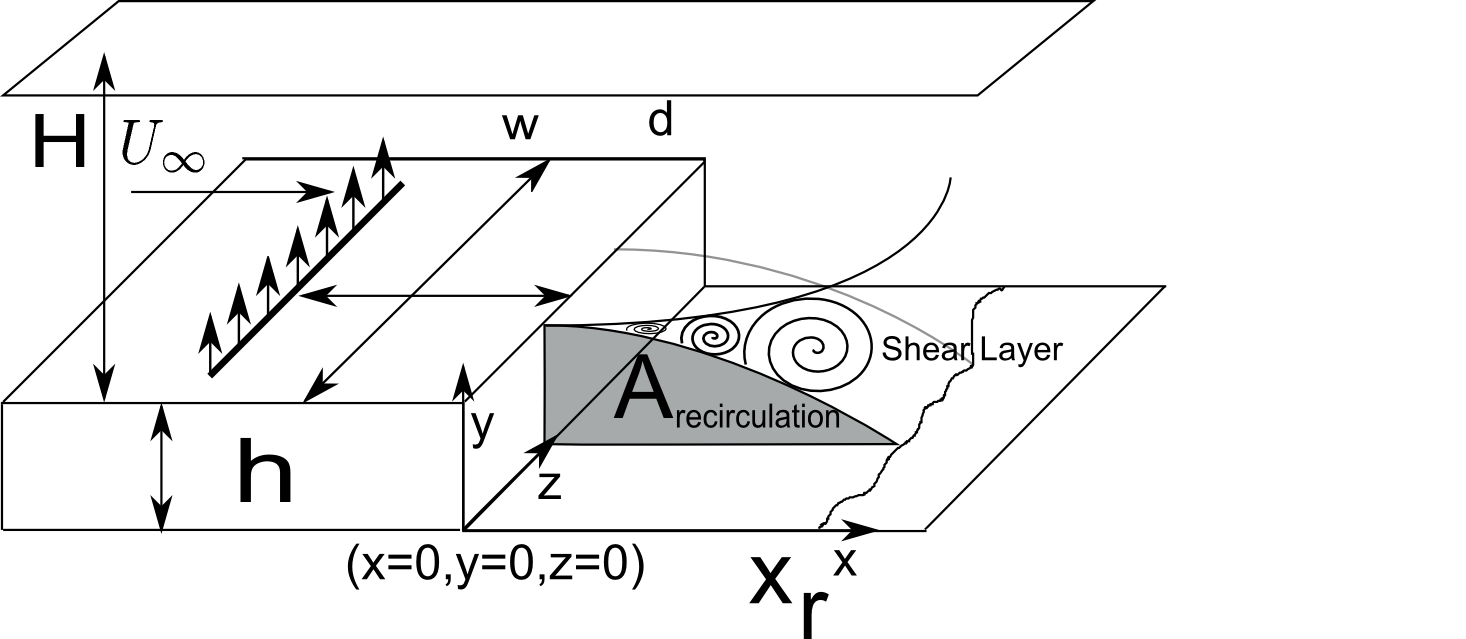}\label{fig:dimensions}}
\subfloat[Sketch of the acquistion apparatus]{\includegraphics[width=0.45\textwidth]{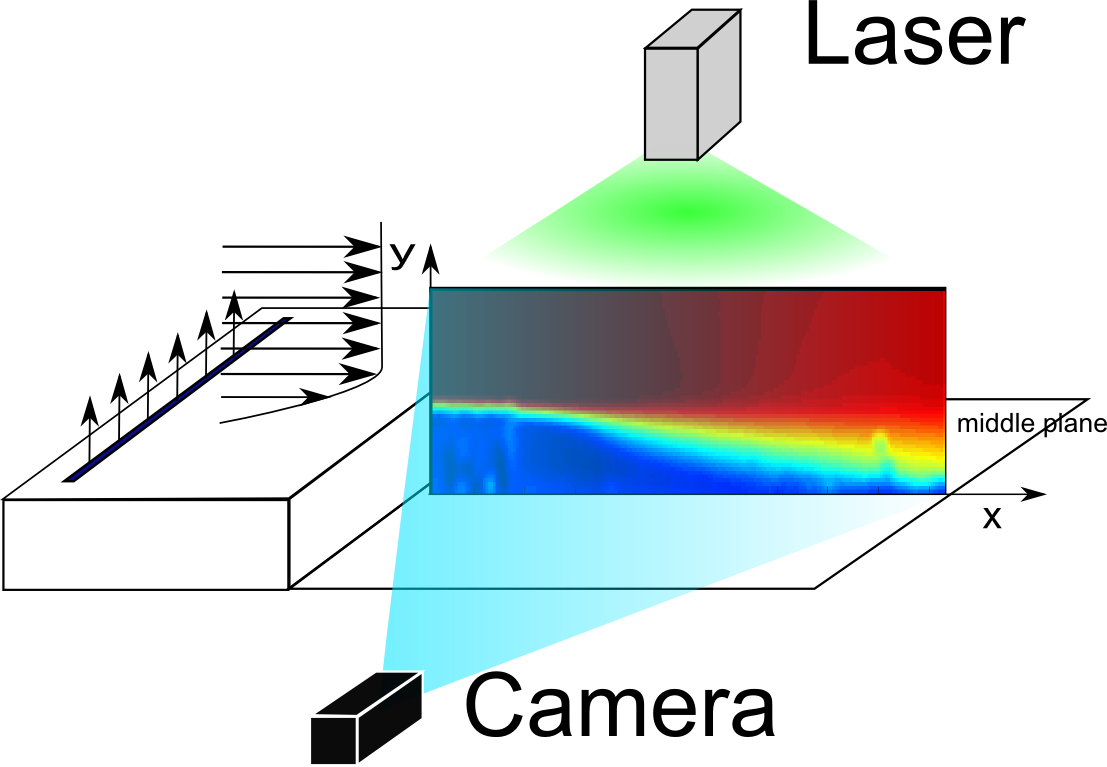}\label{fig:sketch_BFS}}
\caption{}
\end{figure}

The Backward-Facing Step  (BFS)  flow is a benchmark problem, and is commonly used to study massively separated flows both numerically and experimentally (see \cite{Armaly1983,Le1997,JLA2004,JLA2007}). The main features of the BFS flow are the creation of a recirculation downstream of the step together with a strong shear layer in which Kelvin-Helmholtz instability can trigger the creation of spanwise vortices (Figure~\ref{fig:dimensions}). Because separation of the boundary layer is imposed by the step edge, flow control strategies are also limited: it is not possible to delay or trigger the flow separation but only to force the shear layer in a different state to modify the overall recirculation and location of reattachment point \cite{Darabi2004}. 

There are many ways of controlling separated flows as detailed by Fiedler \& Fernholz \cite{Fernholz1990}. Both passive and active actuation methods have been the subject of much research \cite{Chun1996,Mazur2007}. Active actuations using either pulsed or synthetic jets are always located at the step edge in order to ensure maximum effect on the shear layer. Furthermore this is only possible for geometries where the separation line is well-defined, which is not the case for rounded walls or ramps. While this is effective it does burden any setup with additional engineering constraints. For academic purposes this is of lesser concern, while for industrial purposes the cost of spatially imposed actuation can be prohibitive. Furthermore it has been shown that for geometries where separation point can move because of changing external conditions the effectiveness of flow control can be lowered because actuation is no longer where the flow is most receptive \cite{Narayan2002}. Therefore knowing where actuation can be placed upstream while retaining effectiveness is of particular interest. Because of its ability to excite instabilities in the shear layer, pulsed actuation is most efficient when controlling separated flows, as has been shown by \cite{Chun1996,Closkey2002,Breidenthal2001} and used by \cite{Pastoor2008,King2007,Chun1996}. Pulsed jets actuations are defined by several parameters. However while the influence of jet  amplitude and frequency are always modified, signal shape and duty cycle are seldom investigated.
\\\\

An important step in most of closed-loop control strategies is choosing one or several control parameters. The parameter should be either directly computable from sensor data, such as local pressure or drag measurement, or obtained by combining sensor data and a model. The model for closed-loop actuation can be simple (\cite{King2007} recover recirculation length via its correlation to pressure fluctuations) or complicated (\cite{Sipp2010} recovers an approximation of the flow state through Kalman filtering). In the case of the BFS flow, sensors are most often pressure or skin friction sensors, and the control variable is usually the recirculation length $X_r$ (see figure \ref{fig:dimensions}). Wall based sensors present the advantage of high frequency acquisition however they give a limited view of the flow: many phenomena are difficult to access because buried in noise or simply unobservable because vortices in the flow are not visible by the wall sensors. 
\\
Velocity fields can be analyzed to yield a recirculation \emph{area} instead of a length, as show by \cite{Gautier2013control}. It is a measure of how much recirculation is present in a 2D slice of the flow. While the behavior of the recirculation area is often similar to the recirculation length it is not always the case. Because more information about the flow is used to compute the recirculation area it makes sense to use such a variable when possible.
\\
In this paper we investigate the effect of an upstream pulsed jet on the recirculation area downstream of a BFS. The flow state is characterized in the middle plane using real-time optical flow measurements. The parametric space formed by jet amplitude, frequency, and duty cycle is explored for two Reynolds numbers.

\section{Experimental Setup}
\subsection{Water tunnel}
Experiments were carried out in a hydrodynamic channel in which the flow is driven by gravity.
The flow is stabilized by divergent and convergent sections separated by honeycombs. The test section is $80$~cm long with a rectangular cross section $15$~cm wide and $10$~cm high.\\
 The quality of the main stream can be quantified in terms of flow uniformity and turbulence intensity.  The standard deviation $\sigma$ is computed for the highest free stream velocity featured in our experimental set-up. We obtain $\sigma = 0.059$~cm.s$^{-1}$ which corresponds to turbulence levels of $\frac{\sigma}{U_{\infty}}=0.0023$.
 \\
 The mean free stream velocity $U_{\infty}$ can go up to $22$~cm.s$^{-1}$. The Reynolds number is based on the step height h, $Re_h=\frac{U_{\infty} h}{\nu}$, $\nu$ being the kinematic viscosity. A specific leading-edge profile is used to smoothly start the boundary layer which then grows downstream along the flat plate, before reaching the edge of the step $33.5$~cm downstream. The boundary layer is laminar and follows a Blasius profile. The boundary layer thickness  is $\delta = 0.75$~cm for $Re_h=2070$ and $\delta = 0.89$~cm for $Re_h=2900$.

\subsection{Backward-facing step geometry}
The backward-facing step geometry and the main geometric parameters are shown in figure~\ref{fig:dimensions}. The height of the BFS is $h=1.5$~cm. Channel height is $H=7$~cm for a channel width $w=15$~cm. One can define the vertical expansion ratio  $A_y = \frac{H}{h+H} = 0.82$ and the spanwise aspect ratio $A_z=\frac{w}{h+H}=1.76$. 

\subsection{Velocity fields computation}
The flow is seeded with 20~$\mu m$  neutrally buoyant polyamid seeding particles.  The vertical symmetry plane of the test section is illuminated by a laser sheet created by a 2W continuous CW laser beam operating at wavelength $\lambda = 532$~nm passing through a cylindrical lens (Figure~\ref{fig:sketch_BFS}).  The pictures of the illuminated particles are recorded using a Basler acA 2000-340km 8bit CMOS camera. The camera is controlled by a camera-link NI PCIe 1433 frame grabber. Velocity field computations are run in real-time on athe GPU of a Gforce GTX 580 graphics card. \\

The two components of the planar velocity fields (U, V being respectively the streamwise and vertical components) are computed in real-time using an optical flow algorithm \cite{Lucas1984}. Its offline accuracy has been demonstrated by \cite{Plyer2011}. Although there are differences with classic PIV algorithms output velocity field resolution is still tied to the size of the interrogation window. However the output field is dense (one vector per pixel) giving better results in the vicinity of edges and obstacles, which is crucial in BFS flows. Furthermore this gives exceptionally smooth fields. The algorithm was used by \cite{Leclaire2012, Bur2012,Gautier2013control}.

\subsection{Relationship between recirculation length and area.}
\begin{figure}[H]
\centering
\subfloat[Longitudinal velocity field, $Re_h=2900$, no actuation]{\includegraphics[width=0.4\textwidth]{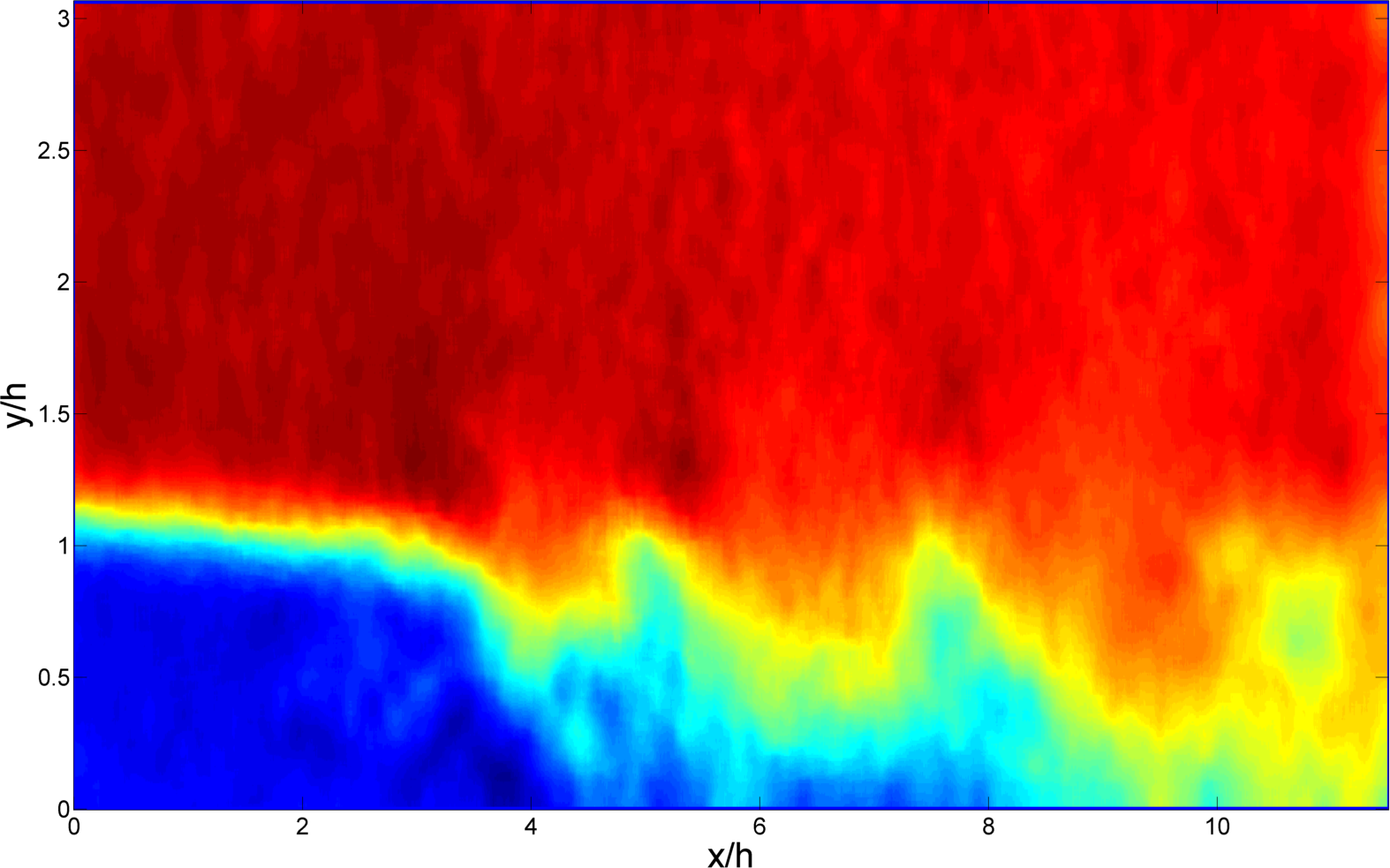}\label{fig:rs}} 
\subfloat[Corresponding instantaneous recirculation area in black using equation \ref{eq:surf}.]{\includegraphics[width=0.4\textwidth]{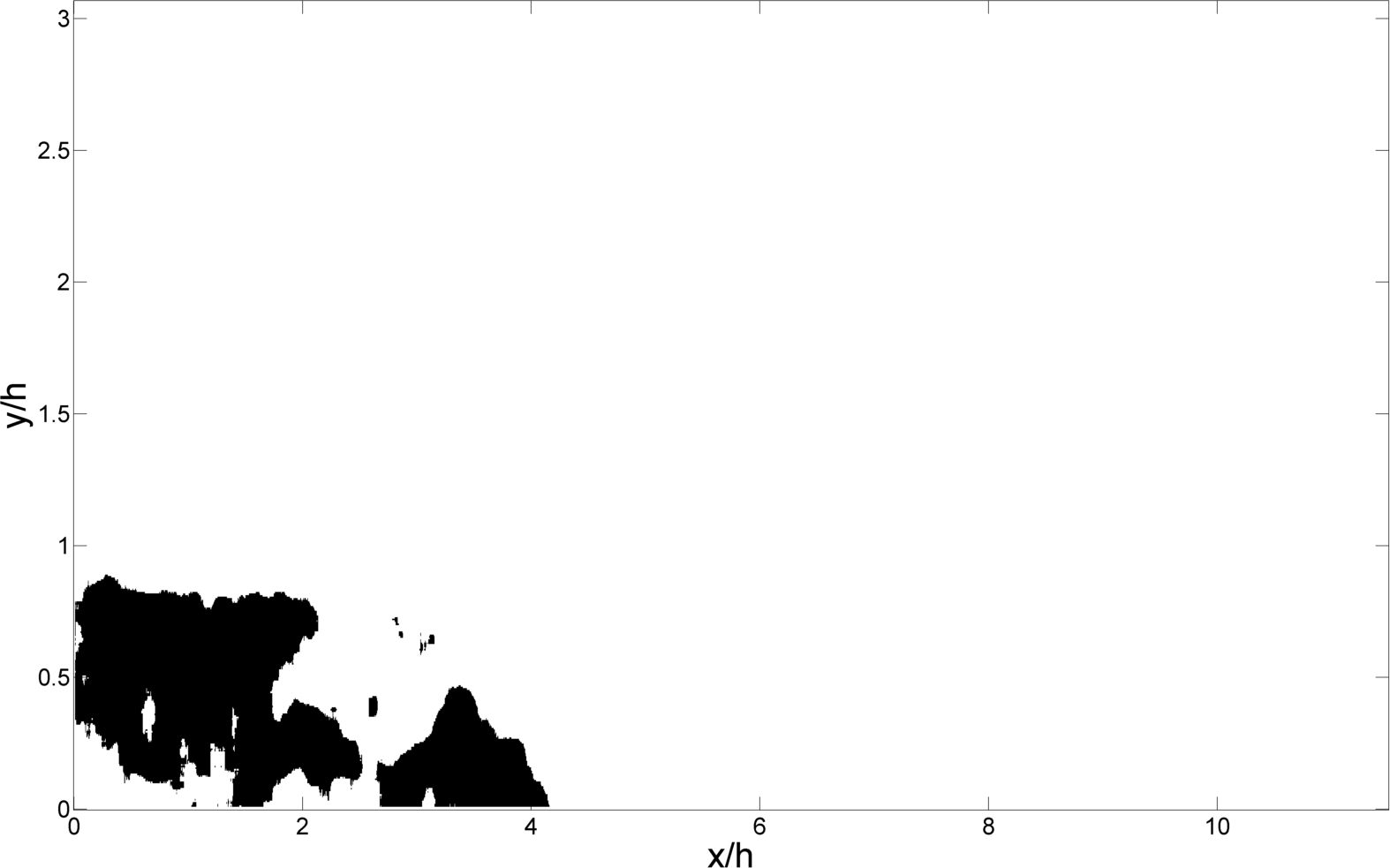}\label{fig:rsb}}
\caption{}
\end{figure}

In the case of separated flows, specifically backward-facing step flows, the length of the recirculation  $X_r$ is commonly used as input variable (\cite{King2007,Chun1996}). There are many ways of computing the recirculation length, however they all give qualitatively similar results \cite{Armaly1983}. Because 2D two-components velocity fields are measured, the recirculation area can be characterized by its area instead of its length \cite{Gautier2013control}.  Building upon 1D definitions the recirculation area can be considered to be the area occupied by the region(s) of flow where longitudinal velocity is negative. The instantaneous recirculation area $A_{rec}$ is then defined in equation \ref{eq:surf}:

\begin{equation}
A_{rec}(t)=\int_{A} H(-v_x) da
\label{eq:surf}
\end{equation}

where $H$ is the Heavyside function. Figures \ref{fig:rs} and \ref{fig:rsb}  show an example of an instantaneous recirculation area. In the following we will consider the mean recirculation area i.e. $A_{rec}$ is computed for every time step for each instantaneous velocity fields before being averaged. It should be noted that it is different from the recirculation area of the mean velocity field.

Because recirculation area is computed from 2D data it has the potential to give a more accurate measure of recirculation in the flow than recirculation length.

It has been shown by Gautier \& Aider \cite{Gautier2013control} that the recirculation area behaves similarly to the recirculation length for varying Reynolds numbers. Moreover previous studies \cite{Armaly1983} have shown the evolution of the recirculation length as a function of the Reynolds number reaches a maximum  between $600<Re_h<1000$ before reaching its asymptotic value for $Re_h > 2000$. The Reynolds numbers featured in this study are high enough to ensure recirculation area has reached its asymptotic regime where recirculation length no longer depends on Reynolds number. For each flow configuration the recirculation area was computed and recorded over 5 minutes with a sampling frequency  $f_a =70$ Hz to ensure convergence. The time series is then averaged over time. It should be noted the recirculation area is computed in real-time, concurrently with image acquisition, therefore only the recirculation area is saved. It avoids saving images and velocity fields, making experimental data very light and greatly hastening the data processing and analysis.

\subsection{Actuation}
\begin{figure}[H]
\centering
\includegraphics[width=0.35\textwidth]{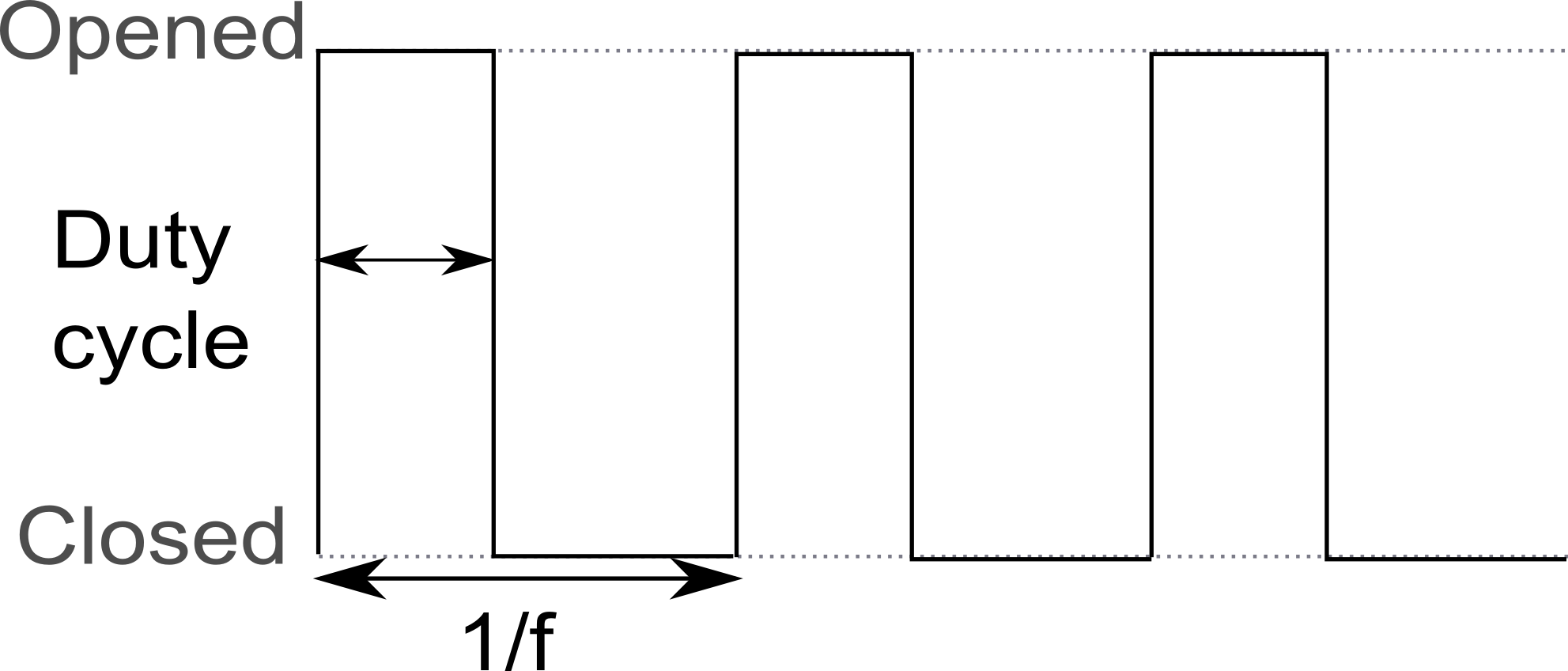}
\caption{Square wave signal and definition of duty-cycle}
\label{fig:square_wave}
\end{figure}

Actuation is provided by a flush slot jet,  0.1~cm long and 9~cm wide. Injection is normal to the wall. The slot is located at a distance $d=3.5$~cm~$= 2.11h$ upstream the step edge (figure~\ref{fig:dimensions}). Water coming from a pressurized tank enters a plenum and goes through a volume of glass beads designed to homogenize the incoming flow. Jet amplitude is controlled by changing tank pressure. The injection geometry was chosen to keep the perturbation as bi-dimensional as possible. 
\\
The flow is modulated by a one-way voltage driven solenoid-valve. It is controlled by a square-wave signal described in figure \ref{fig:square_wave} with an actuation frequency $f_a$. The square wave signal was chosen for its simplicity,  other signal forms could be considered.  

The duty-cycle $dc$ (in \%) is the ratio between the time for which the valve is opened over time of a cycle. Jet amplitude is defined as the ratio between mean jet exit velocity when the jet is active and cross flow velocity $a_0=\frac{U_{jet}}{U_0}$. The duty cycle therefore has no impact on jet amplitude.

\subsection{Natural shedding frequency}
Kelvin-Helmholtz instabilities in the shear layer create spanwise vortices which in turn influence the recirculation area. 
An effective way of detecting such vortices is to compute on the two-components 2D velocity fields the swirling strength criterion $\lambda_{ci}(s^{-1})$. It was first introduced by \cite{Chong1990} who analyzed the velocity gradient tensor and proposed that the vortex core be defined as a region where $\nabla \bf{u}$ has complex conjugate eigenvalues. For 2D data we have  $\lambda_{Ci}=\frac{1}{2}\sqrt{4 \det(\nabla \bf{u})- \Tr(\nabla \bf{u})^2}$ when such a quantity is real, else $\lambda_{Ci}=0$.   It was later improved and used for the identification of vortices in three-dimensional flows by \cite{Zhou1999}. 

\begin{figure}[H]
\centering
\includegraphics[width=0.45\textwidth]{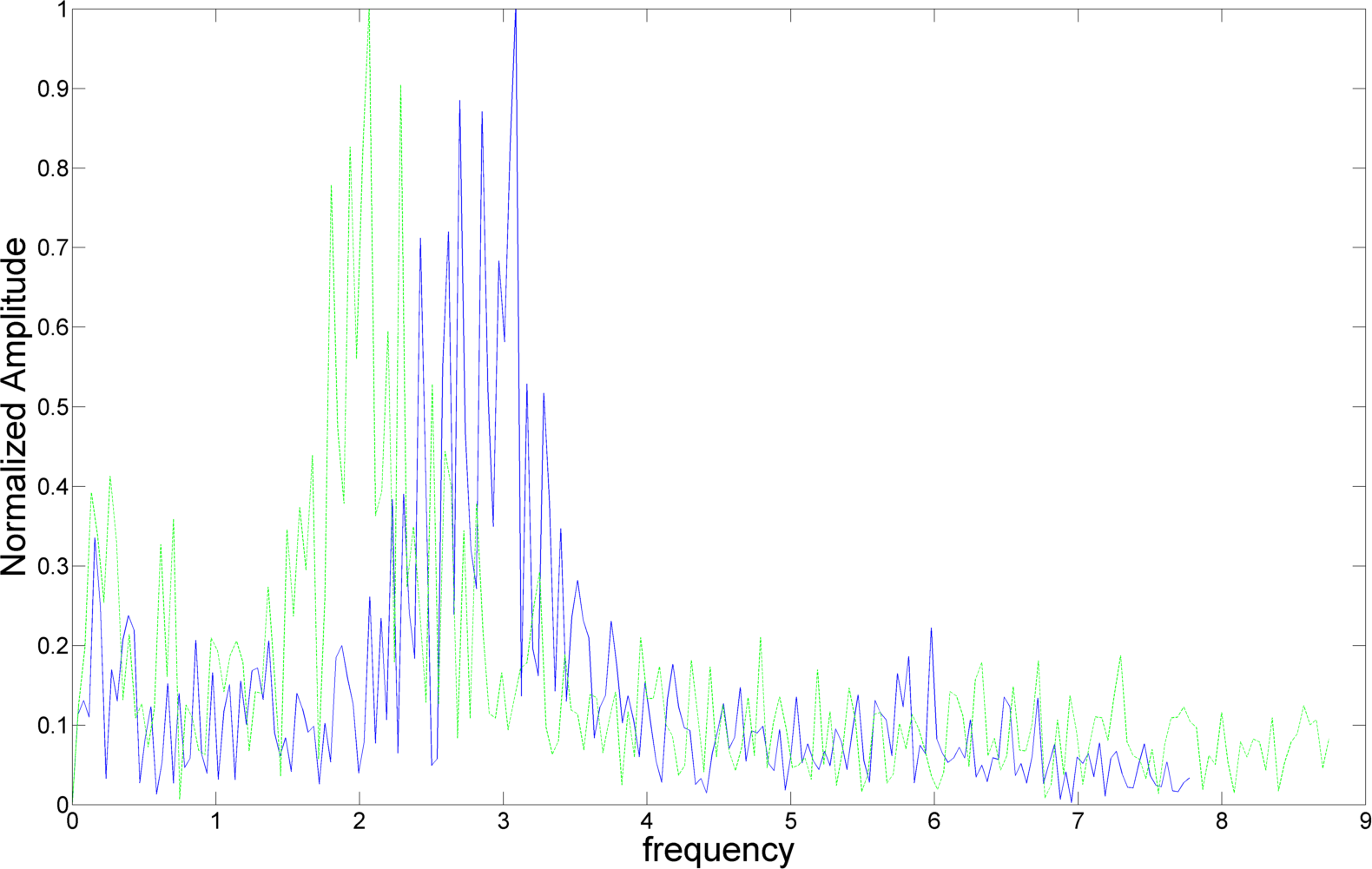}
\caption{Frequency spectrum for $Re_h=2070$ (peak at $St_h=0.258$) in dashed green and for $Re_h=2900$ (peak at $St_h=0.272$) in solid blue}
\label{fig:freq_spectrum}
\end{figure}

The shedding frequency is obtained by spatially averaging $\lambda_{Ci}$ in the vertical direction at $x=3h$ with a sampling frequency $f_s = 40$Hz. Essentially vortices are counted as they pass through an imaginary line. Figure \ref{fig:freq_spectrum} shows frequency spectra obtained by Fourier transform for both Reynolds numbers, where $St_h=\frac{f h}{U_0}$ is the Strouhal number based on the step height.

\section{Results}

\subsection{Influence of frequency}
\begin{figure}[H]
\centering
\includegraphics[width=0.45\textwidth]{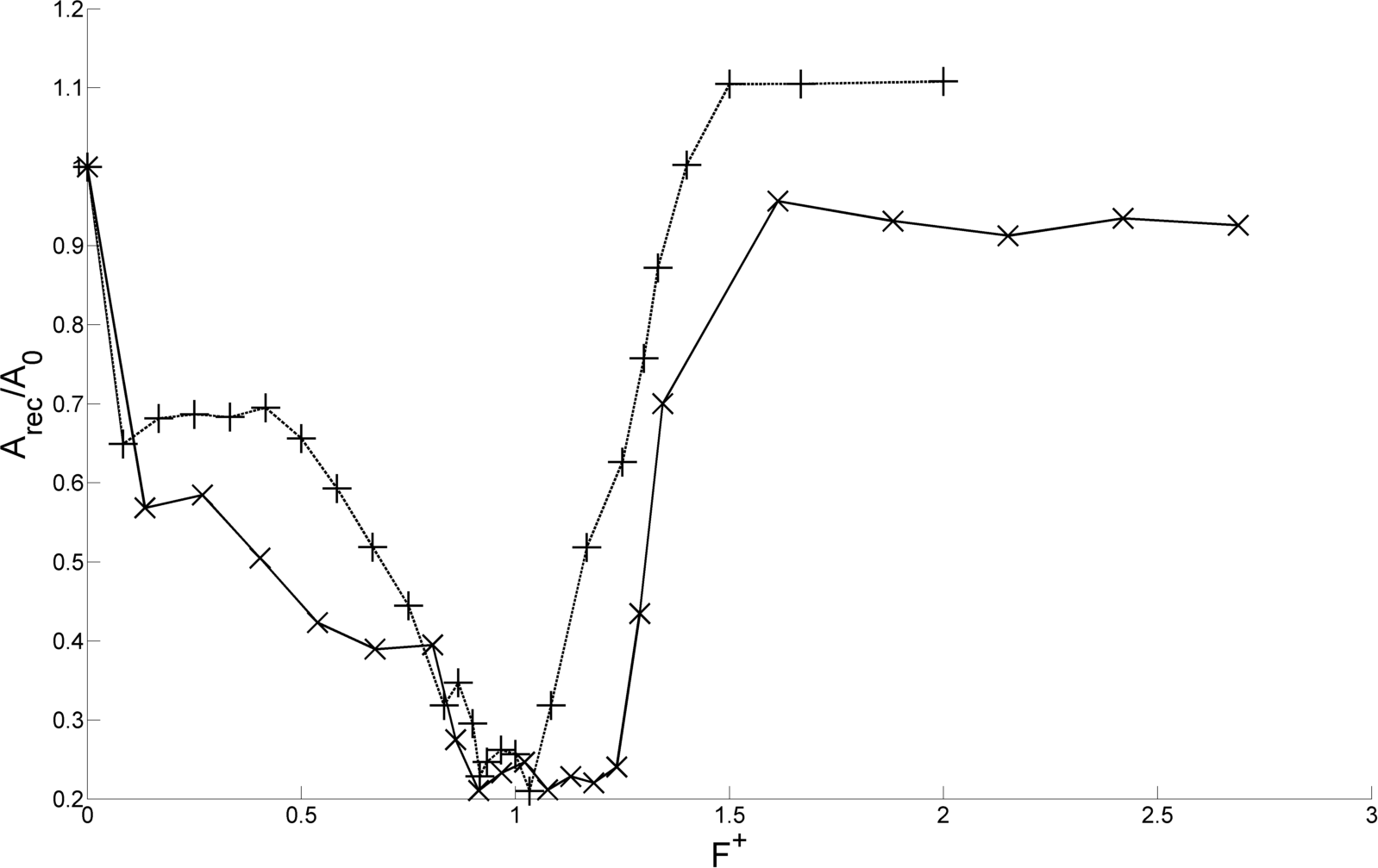}
\caption{Evolution of time averaged recirculation area $\frac{A_{rec}}{A_0}$ as a function of the frequency for $Re_h=2070$ ($\times$) and $Re_h=2900$(+) with $dc =$ 50\% and $a_0=0.040$.}
\label{fig:recirc_freq}
\end{figure}

Figure \ref{fig:recirc_freq} shows the evolution of recirculation area (non-dimensionalized by the uncontrolled recirculation area $A_0$) when frequency varies for both Reynolds numbers. Jet amplitude and duty cycle are kept constant. Jet amplitudes were chosen empirically. Previous open-loop control experiments have shown reduction in circulation length of up to 40 \% \cite{Chun1996}. Here recirculation area is decreased by as much as 80 \%. The reduction is maximum when the pulsing frequency is close to vortex shedding frequency, $f\simeq f_0$, i.e. $F^+ = \frac{f_a}{f_0}  \approx 1$. This result is similar to the effect of flow control at the step edge.

These results show how upstream actuation can effectively control a backward-facing step flow. The influence of the upstream location of the actuator was beyond the scope of the present study.

\begin{figure}[H]
\centering
\includegraphics[width=0.45\textwidth]{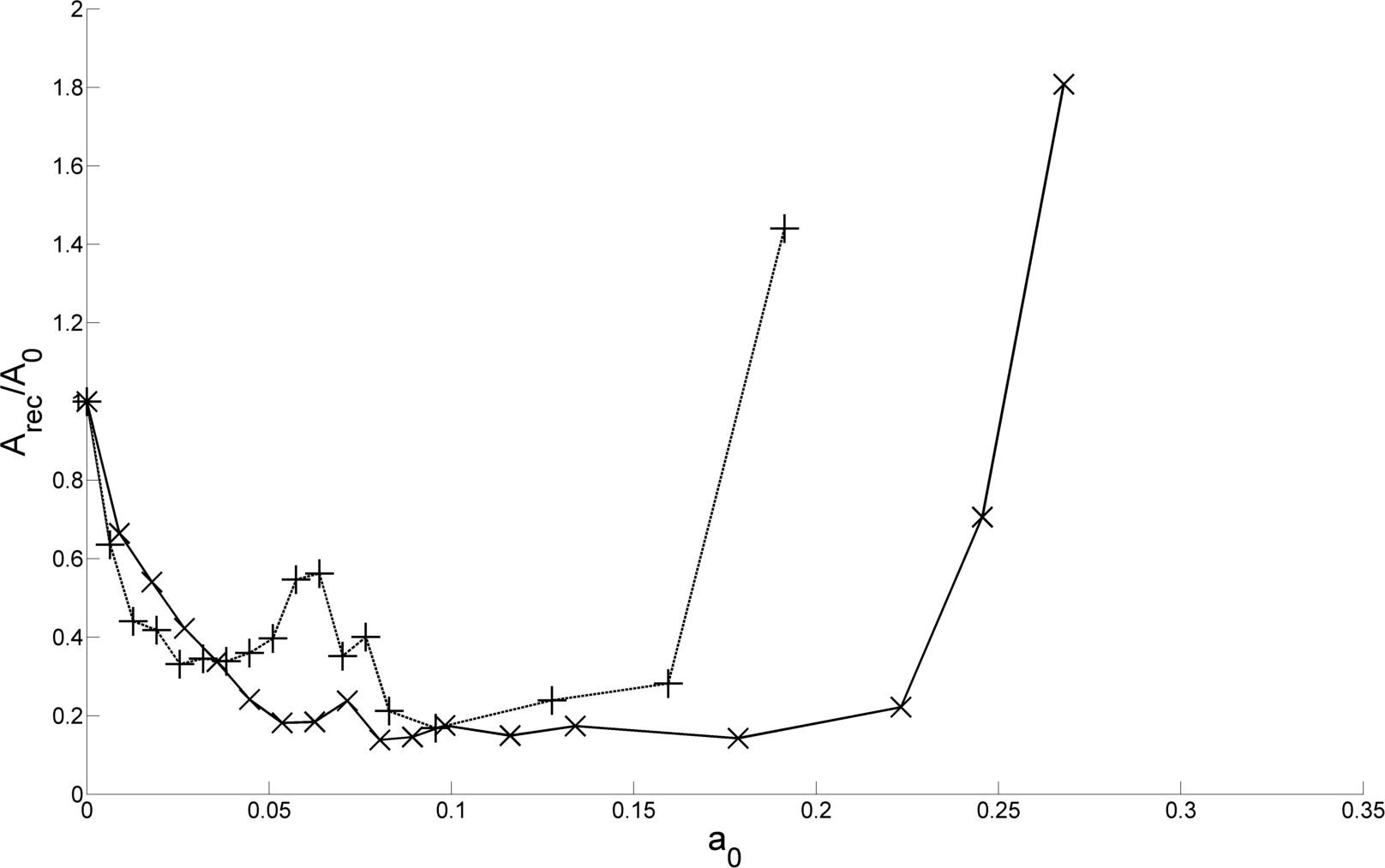}
\caption{Evolution of time averaged recirculation area $\frac{A_{rec}}{A_0}$ as a function of $a_0$ for $Re_h=2070$($\times$) and $Re_h=2900$(+) with $F^+   \approx 1$ and $dc=50\%$}
\label{fig:recirc_P}
\end{figure}

\subsection{Influence of jet exit velocity}
Figure \ref{fig:recirc_P} shows the evolution of recirculation area when jet amplitude varies, for both Reynolds numbers. The actuation frequency giving maximum reduction was chosen for both Reynolds numbers ($F^+   \approx 1$) and duty cycle was kept constant at $dc = 50$\%. One can clearly see that there is an optimal amplitude for the jet: if too small or too large, the control looses its efficiency, the optimum ratio being around $a_0 \approx 0.1$. In this case, the reduction of the recirculation area is even larger, close to 85\%.

These results highlight the main difference between edge and upstream jet actuation. Similarly to edge injection, a minimal jet amplitude is required to affect the flow. However for upstream actuation, recirculation area increases with increasing jet amplitude instead of decreasing. Indeed, for high amplitudes the jet fully penetrates the cross-flow effectively becoming an obstacle to the incoming flow, leading to a massive increase of the recirculation area. One also notes that flow behavior is similar for both Reynolds numbers. Once again while the recirculation area of the controlled mean field is near 0 \% mean recirculation area is closer to 10 \% of the uncontrolled values.

\subsection{Influence of duty cycle}
\begin{figure}[H]
\centering
\includegraphics[width=0.45\textwidth]{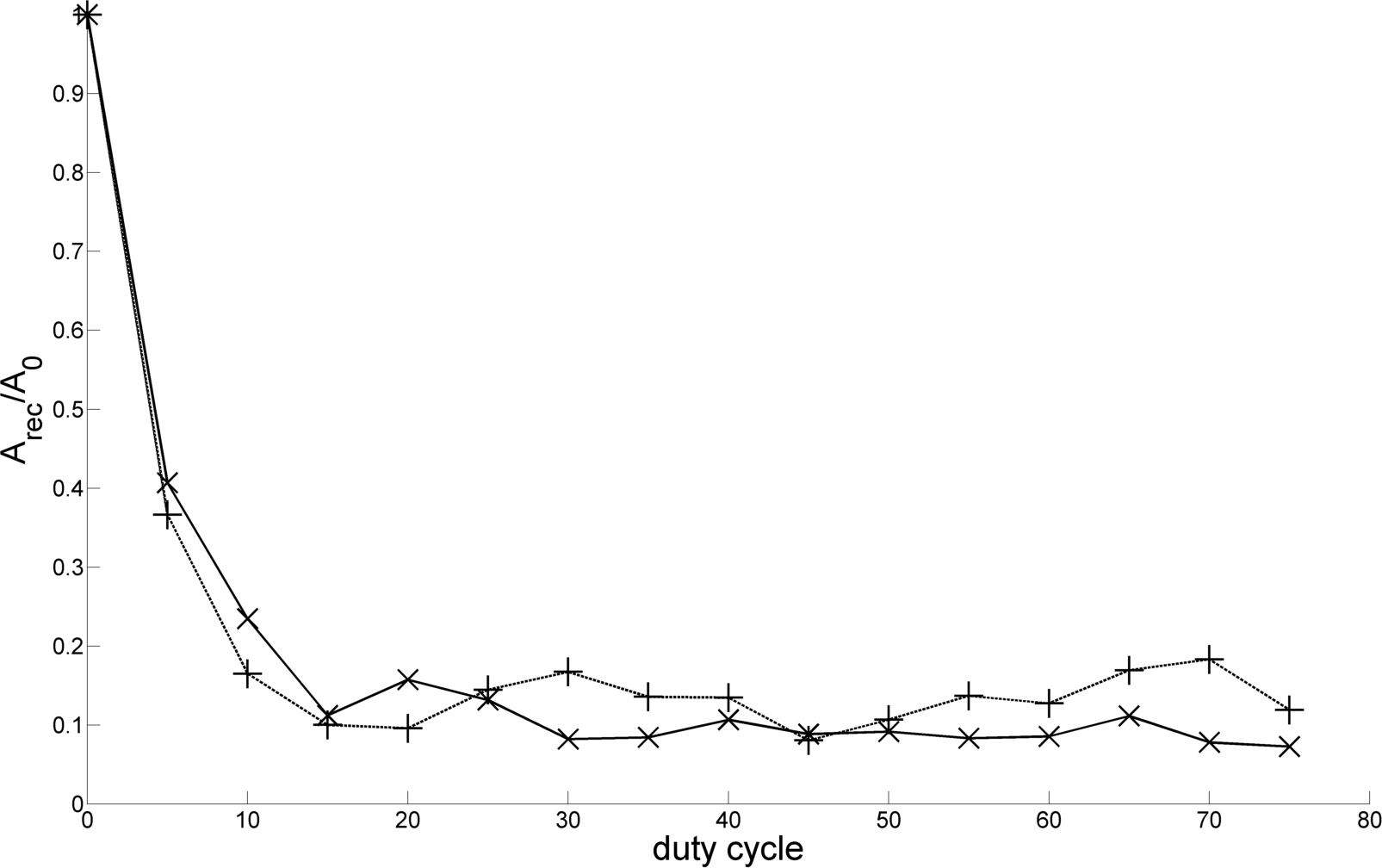}
\caption{Time averaged recirculation area as a function of the duty cycle for $Re_h=2070$ ($\times$) and $Re_h=2900$ (+)  with $F^+   \approx 1$ and $a_0=0.083$}
\label{fig:recirc_dc}
\end{figure}

Figure \ref{fig:recirc_dc} shows the evolution of the recirculation area as a function of the duty-cycle for both Reynolds numbers and for the optimal actuation frequency and amplitude previously found. 

Recirculation reduction area is increased, reaching nearly 90\%. A minimal duty cycle of 10 \% is required to fully affect the flow, much lower than the usual 50 \% used in most of the previous studies. This is an important result:  the duty cycle can be brought down significantly while still maintaining an effective control and then allowing a strong improvement of the overall energy balance between power used by the actuation and power gain (if related to a drag decrease, for instance). 

\subsection{Recirculation suppression}
\begin{figure}[H]
\centering
\subfloat[Comparison of time averaged recirculation area obtained for $Re_h=2070$ for the uncontrolled (grey) and controlled (black) configuration ($F^+   \approx 1$, $dc=20$, $a_0=0.083$). ]{\includegraphics[width=0.45\textwidth]{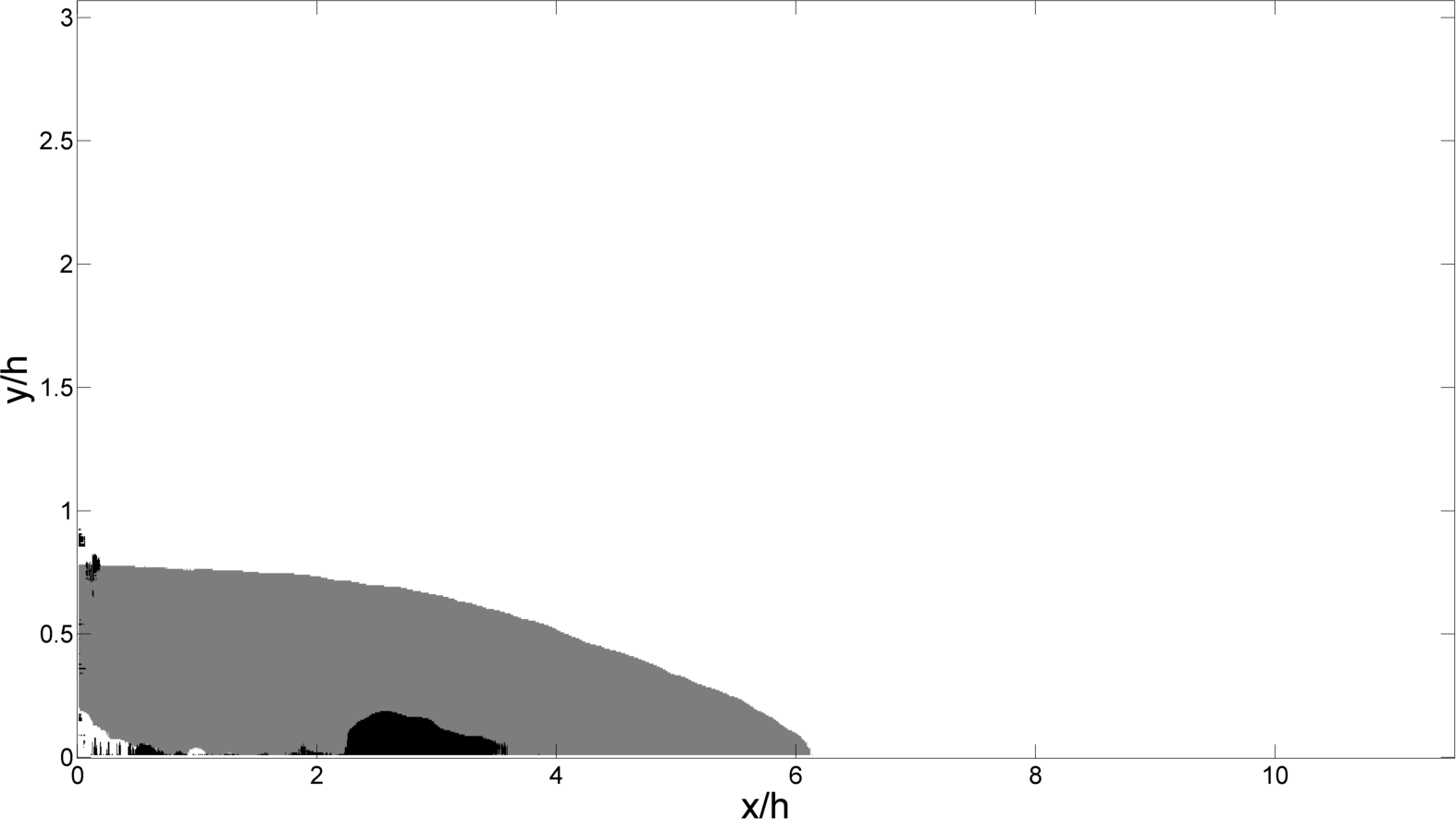}\label{fig:opt_Re1}} 
\hspace{5mm}
\subfloat[Comparison of time averaged recirculation area obtained for $Re_h=2900$ for the uncontrolled (grey) and controlled (black) configuration  ($F^+   \approx 1$, $dc=20$, $a_0=0.088$).]{\includegraphics[width=0.45\textwidth]{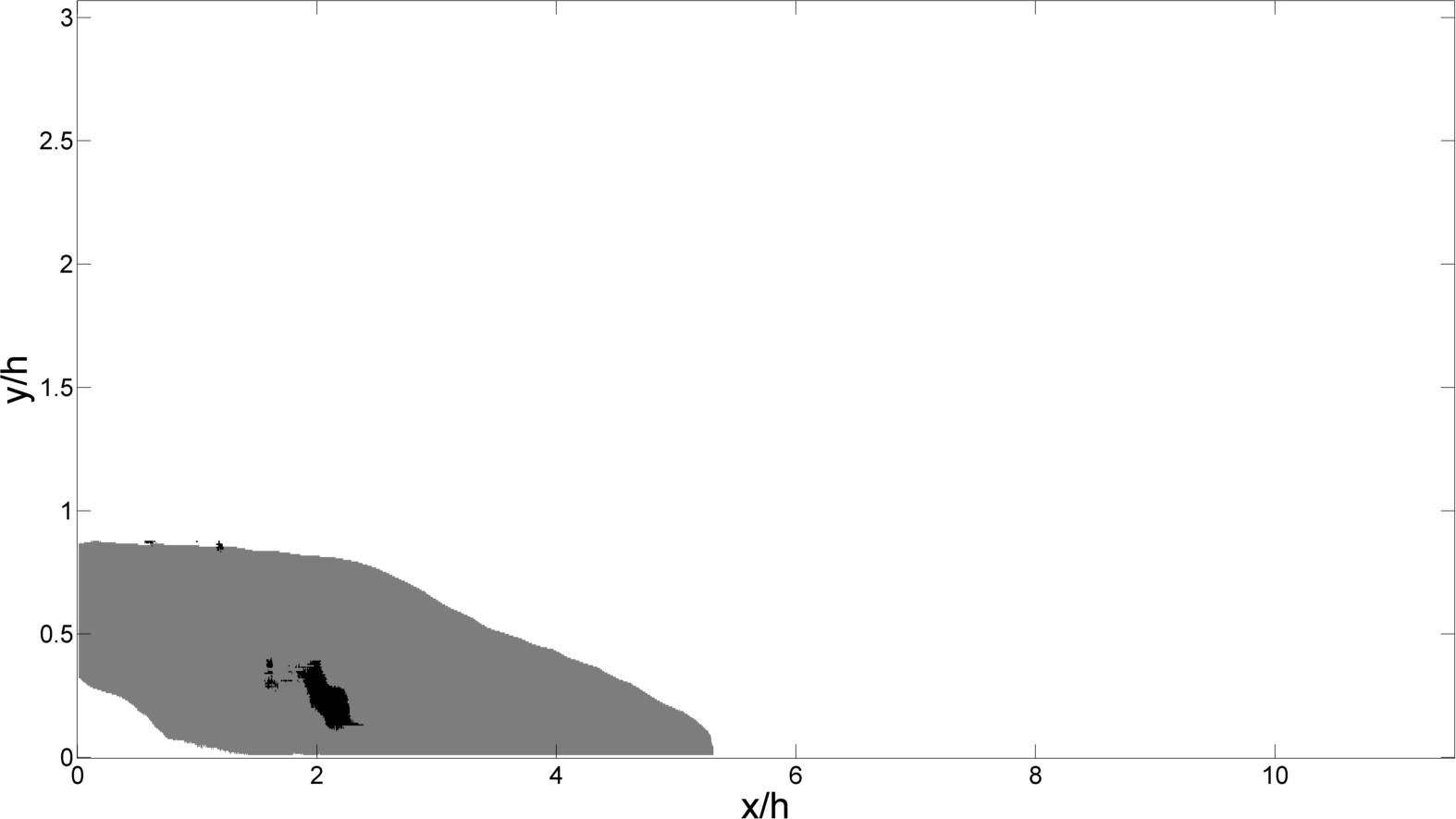}\label{fig:opt_Re2}}
\caption{}
\end{figure}

It appears one major difference between recirculation area and recirculation length is its sensitivity to actuation. While the recirculation length can be reduced by 40 \%, the recirculation area can be reduced by nearly 90 \%.
To explain this, velocity fields were computed at optimal parameters for both Reynolds numbers. Figures \ref{fig:opt_Re1} and \ref{fig:opt_Re2} show the recirculation area of the mean field in the uncontrolled and optimally controlled cases. On average there is very little recirculation in the controlled cases. Indeed, recirculation area can be almost null while recirculation length remains significant.

These figures illustrate how, in the mean sense, recirculation can be cancelled through targeted control. Presumably the same results could be obtained using a pulsed actuation at the step edge, but it has to be confirmed.

\section{Conclusion}
The flow downstream of a backward-facing step controlled by an upstream pulsed jet was experimentally studied in a hydrodynamic channel. A measure of the recirculation area downstream of the BFS was introduced and used to quantify the effect of actuation for several flow configurations. The parametric space formed by jet amplitude, actuation frequency and duty cycle was explored for Reynolds numbers $Re_h=2070$ and $Re_h=2900$. 
\\
Results show recirculation can be greatly reduced, and in some cases nearly suppressed, for a fairly wide ranges of actuation parameter. Furthermore while this phenomenon is clearly observed when considering recirculation area, it can be missed when considering only recirculation length.  It emphasizes the importance of properly choosing the criterion used to evaluate the state of the separated flow.  Recirculation area gives a more global evaluation of the state of the flow than recirculation length.
\\
The investigation of jet amplitude shows that, in the same way as injection at the step edge, a minimal jet amplitude is required to better control the flow. However step injection and upstream injection differ at high jet amplitudes. In contrast to step injection where it has been shown that raising jet amplitude merely increases actuation effectiveness, albeit with diminishing returns, upstream injection jet amplitude reaches a threshold above which recirculation is greatly increased instead of decreased. In the case of upstream actuation, an optimal jet amplitude can be found.
\\
Finally actuation is shown to be effective over a wide range of duty cycles, reaching a reduction of the recirculation area close to 90 \%.  Moreover, it is shown that the duty cycle can be lowered to 10 \% while keeping recirculation at a minimum. Furthermore it is likely this limit is a consequence of the imperfect nature of the actuator. A better actuator could achieve lower duty cycles. Thus performances can be maintained while considerably lowering flow rate injection, and therefore energy expenditure. This result is of great interest for flow control applications where energy balance is a crucial point.

\section*{Acknowledgement}
The DGA (Direction G\'en\'erale de l'Armement)  is gratefully acknowledge for its financial support.

\bibliographystyle{unsrt}	
\bibliography{Bibliography}

\end{document}